\documentclass{article}

\usepackage{arxiv}

\usepackage[utf8]{inputenc} \usepackage[T1]{fontenc}

\PassOptionsToPackage{hyphens}{url}

\usepackage[breaklinks,backref=page]{hyperref}

\usepackage{url}
\usepackage{booktabs}
\usepackage{amsfonts}
\usepackage{nicefrac}
\usepackage{microtype}
\usepackage{lipsum}
\usepackage{graphicx}
\usepackage[numbers,sort&compress]{natbib}
\usepackage{doi}
\usepackage{float}
\usepackage{multirow}
\usepackage{lineno}

\usepackage{cleveref}

\usepackage{wrapfig}

\usepackage{authblk}

\usepackage[justification=centering]{caption}

\usepackage[inline]{enumitem}

\usepackage{bbding}

\usepackage[flushleft]{threeparttable}

\usepackage{listings}

\usepackage{subcaption}

\renewcommand*{\backref}[1]{}
\renewcommand*{\backrefalt}[4]{
  \ifcase #1
    No citations.\or
(Cited on page: #4).
  \else
(Cited on pages: #4).
  \fi
}

\title{ESPORT: Electronic Sports Professionals Observations and Reflections on Training }

\date{\today}

\makeatletter
\newcommand\email[2][]{\newaffiltrue\let\AB@blk@and\AB@pand
      \if\relax#1\relax\def\AB@note{\AB@thenote}\else\def\AB@note{\relax}
        \setcounter{Maxaffil}{0}\fi
      \begingroup
        \let\protect\@unexpandable@protect
        \def\thanks{\protect\thanks}\def\footnote{\protect\footnote}
        \@temptokena=\expandafter{\AB@authors}
        {\def\\{\protect\\\protect\Affilfont}\xdef\AB@temp{#2}}
         \xdef\AB@authors{\the\@temptokena\AB@las\AB@au@str
         \protect\\[\affilsep]\protect\Affilfont\AB@temp}
         \gdef\AB@las{}\gdef\AB@au@str{}
        {\def\\{, \ignorespaces}\xdef\AB@temp{#2}}
        \@temptokena=\expandafter{\AB@affillist}
        \xdef\AB@affillist{\the\@temptokena \AB@affilsep
          \AB@affilnote{}\protect\Affilfont\AB@temp}
      \endgroup
       \let\AB@affilsep\AB@affilsepx
}
\makeatother

\author[1]{\textbf{Andrzej Białecki}\textsuperscript{*,}} 
\affil[1]{Warsaw University of Technology}

\author[2]{\textbf{Peter Xenopoulos}}
\affil[2]{Independent Researcher}

\author[3]{\textbf{Paweł Dobrowolski}}
\affil[3]{Institute of Psychology, Polish Academy of Sciences}

\author[4]{\textbf{Robert Białecki}}
\author[4]{\textbf{Jan Gajewski}}
\affil[4]{Józef Piłsudski Warsaw University of Physical Education}

\hypersetup{
pdftitle={ESPORT Electronic Sports Professionals Observations and Reflections on Training},
pdfauthor={Andrzej Bialecki},
pdfkeywords={esport, human-computer interaction, training, interviews, perception},
}

\begin{document}
\maketitle

\let\thefootnote\relax\footnote{\textsuperscript{*} Corresponding author: \url{andrzej.bialecki94@gmail.com}}
\let\thefootnote\relax\footnote{\textsuperscript{*} Institutional contact: \url{andrzej.bialecki.dokt@pw.edu.pl}}

\begin{abstract}

    Esports and high performance human-computer interaction are on the forefront of applying new hardware and software technologies in practice. Despite that, there is a paucity of research on how semi-professional and professional championship level players approach aspects of their preparation.

    To address that, we have performed, transcribed, and analyzed interviews with top-tournament players, coaches, and managers across multiple game titles. The interviews range from competitive events occuring between 2015-2020. Initial processing included transcription and manual verification. The pre-processed interview data were then organized and structured into relevant categories, touching on psychological, physical, and nutritional aspects of esports preparation. Further, where applicable, interview responses where rated and quantified via consensus judgement by a panel of experts.

    The results indicate that physical training was most often mentioned as a relevant or consistent activity, while nutrition was indicated as relatively unimportant. Qualitative analysis also indicated that consistency and resiliency were noted as the most key factors recommended for upcoming esports competitors. It is also clear that many players put emphasis on balancing their gameplay time and with activities. Lastly, we identified important areas of inquiry towards a deeper understanding of the mental and physical demands of professional esports players.

\end{abstract}


\section{Introduction}
\label{sec:Introduction}

Esports can be viewed as a subset of gaming that leverages simulated environments to define the competition space. There are numerous efforts in the research community to preciesely define its role in the sports landscape. This includes efforts towards a more precise definition of esports \cite{BialeckiRedefiningSports2022,Fried2023,Freeman2017}. The largest esports tournaments routinely attract millions of concurrent viewers and award millions of dollars in prizes. Thus, esports is comparable in interest with popular conventional sports and is a largely global phenomenon, with tournaments held year-round on every continent \cite{URLPei2019,URLEsportsCharts}. Such strong growth has spawned not only a burgeoning esports industry, but also a plethora of interdisciplinary research, detailing the intersection of esports with law \cite{Holden2017}, health \cite{Rudolf2020,Arnau2023,Madden2021}, and media studies \cite{Hamilton2014,Torres2022,Kow2013}. In particular, much esports-focused work fits well within the domain of human-computer interaction, given the mechanism by which many esports function \cite{Watson2021}. The prominent use of hardware and software in esports means that research in this area could be attributed to human-computer interaction (HCI) \cite{Chiu2021}. It is common to see reports on technological advancements in the area of gaming peripherals, some teams sell their own branded hardware yet their influence in the design process is unknown.

\begin{figure}[H]
    \centering
    \includegraphics[width=0.9\linewidth]{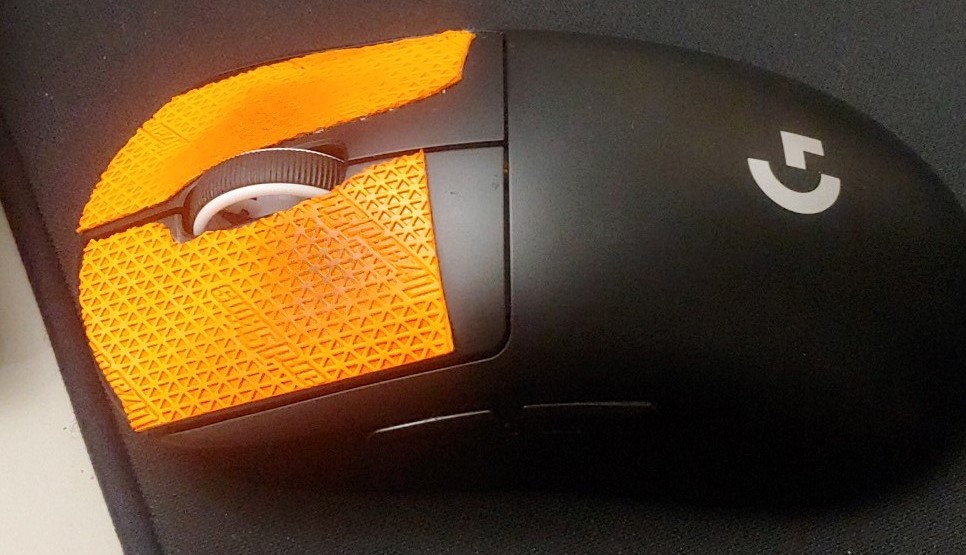}
    \caption{Examples of equipment customized by professional esports players (Source Mikołaj ``Elazer'' Ogonowski). Notice and finger wear marks and taped buttons.}
    \label{fig:CustomizedMouse}
\end{figure}

There are no limits to use of customized pointing devices in most if not all of esports, including adaptive hardware \cite{Tabacof2021}. Players often tinker with their equipment, personalizing it to fit their requirements for comfortable play. We show such customized hardware on \autoref{fig:CustomizedMouse}. Ergonomy and ease of use of hardware and software is a known area in esports. There are multiple ways to position the mouse and keyboard, and different types of mouse grips that players are used to. Other known issue in this space concerns the latency in multiplayer competition, and potential software inconsistencies \cite{Sheldon2003}.

Esports encompasses a large community, including players of all skill levels, fans, and commentators "casters", and has provided intriguing directions for HCI \cite{Kriglstein2021}. This includes cognitive and mental health in esports \cite{Wu2021,Madden2021,Selen2020} and visualization systems for esports \cite{Xenopoulos2022,Feitosa2015,Afonso2019,Kuan2017,Rijnders2022,Wallner2021}. Additionally, computer games offer unique opportunities for cognitive research on neuroplasticity \cite{Kowalczyk2018}, and in their richness of stimuli can be compared to how biofeedback facilitates skill acquisition \cite{Chen2021}. Despite the large-scale community and growing interest that esports provides, there has been little work directed towards the long tail of the distribution of player skill - namely, highly skilled and professional esports players. These players provide a unique unit of study in that their skill often warrants distinct mental, cognitive, and physical demands. At the same time, sports science has repeatedly investigated the aforementioned characteristics for professionals in other sports \cite{Lapi2018,Perrey2022,Walton2018}, yet this seems to be limited for the esports domain despite its cognitive nature. Furthermore, many esports are remarkably different from conventional sports, given their online nature and that they often mandate a sedentary style of play, thus making it challenging to extrapolate conclusions from conventional sports. Existing systematic reviews and meta analyses point towards the need of further research in esports \cite{Kelly2021,Banyai2019}. This includes research on physical activity in esports \cite{Voisin2022}. In this paper, we aim to bridge this gap by focusing on highly-skilled esports players.

Despite growing academic interest in esports, esports curricula, and high performance human-computer interaction, there seems to be a lack of information regarding the needs of esports athletes from first hand sources. Accordingly, we apply both qualitative and quantitative analysis of interviews with esports professionals to investigate their approach in three major categories of physical activity, psychological factors in esports, and nutrition. Uniquely, we draw upon a curated dataset of a highly-skilled subset of players representing the top end of esports players, many of which have longstanding fanbases or storied careers. These interviews were gathered by visiting major esports tournaments such as Intel Extreme Masters Katowice.

The goal of our work is to highlight the perspective of professional esports staff on various aspects of the training process, including some of the top players in the world in different games. We have formulated the following main research questions to facilitate our approach towards our goal:
\begin{itemize}[leftmargin=1.5cm]
    \item[\textbf{RQ 1:}] What is the players' attitude towards physical activity, nutrition, and psychology?
    \item[\textbf{RQ 2:}] Do players have access to support staff that takes care of physical activity, nutrition, and psychology?
    \item[\textbf{RQ 3:}] What is the perceived path to becoming an esports player?
\end{itemize} \section{Related Work}
\label{sec:RelatedWork}

Similar to our attempt here, a few other works have employed interview-based investigations. This includes research on coaching practices and issues; one such study focused on two main categories in a qualitative manner, namely general esports coaching information and challanges in League of Legends (LoL) esports. The findings of this study emphasized different coaching philosophies, ideas of trust, synergy, micro play, macro play, and other psychological components that were perceived as key in a well-functioning team environment \cite{Sabtan2022}.

On the other hand, it is not only coaches who create the team environment. Some investigations focused on the player-perceived determinants of success, mostly in the area of psychology through thematic analysis \cite{Poulus2022}. Furthermore, other authors attempted to quantify the amount of physical activity, and in-game training. They have found that press overexaggerated the training time to be within 12 to 14 hours a day, while the players indicated on average 5.28 hours per day \cite{Kari2016}. Other researchers performed 12 interviews with with 11 professional Overwatch esports players performing a themathic analysis on game-sense and mechanics. They have concluded that important aspects are knowledge of ally and enemy positioning, and timing \cite{Fanfarelli2018}. Similar to our approach there also exists an article on perceptions of effective training in esports, although it focused on League of Legends practitioners. Authors split interview gathered data into 3 themes: the state of training, training experiences, and motivational change. Results indicated that the training approaches and motivation can shift. Within the results, it was pointed out that players often perceive their training practices as sub-optimal. Additionally authors highlighted the emotional and physical toll on the players. This is due to the extensive hours of gameplay in sedentary position to improve ones ability (``grinding'') \cite{Abbott2023}.

A Scoping Review on associations between esports participation and health habits uncovered that most of the players 65\%-88.7\% indicated regular exercise habits and indicated relative novelty of the esports research area \cite{Pereira2022}. Additionally, though different from our attempt, there exist interview based investigations on how esports athletes perceive performance-enhancing substances (doping) \cite{Schubert2022}.

\section{Material and Methods}
\label{sec:MaterialAndMethods}

\subsection{Interview Collection}
\label{sec:interview_collection}

We collected interviews with esport athletes, coaches, and managers ranging from semi-professional to professional. All of the interviews were collected in a form of audio-visual recordings in English (22 interviews) or in Polish (17 interviews), the total number of interviews split by the interviewee type is availabile in \autoref{tab:interview_stats}. Collection of the interviews took place during events such as
\begin{enumerate*}[label=(\arabic*)]
    \item ESL One Cologne 2015,
    \item ESL One Frankfurt 2015,
    \item IEM Katowice 2016, 2017, 2018, 2019
    \item ESL Polish Championships 2017,
    \item WGL EU Season 2 Qualifier 2017,
    \item MeetPoint 2018,
    \item ESL One Katowice 2019.
\end{enumerate*}

All of the interviews were published online or recorded for further broadcast in cable television. Interviewees played, or were related to games such as League of Legends (LoL), StarCraft 2 (SC2), Heroes of the Storm (HoTS), Counter-Strike: Global Offensive (CS:GO), World of Tanks (WoT), Dota 2, and the Business side of esports.

The interviews were originally recorded from 2015 up until 2020 and span multiple game genres and esports titles. We used a total of 39 interviews. The total number of questions asked per type of the interviewee is described in \autoref{tab:interview_stats}.

\begin{table}[H]
    \caption{Interview count breakdown 2015--2020}
    \label{tab:interview_stats}
    \centering
    \begin{tabular}{lcc}
        \hline
        type                & interviews & total questions \\ \hline
        player              & 28         & 296             \\
        coach (in-game)     & 2          & 20              \\
        coach (performance) & 1          & 11              \\
        manager             & 7          & 63              \\
        commentator         & 1          & 6               \\ \hline
        total               & 39         & 396             \\ \hline
    \end{tabular}
\end{table}

\subsection{Data Pre-Processing}
\label{sec:pre_processing}

The interviews were initially processed with OpenAI Whisper, which provides automatic video to text transcription \cite{Radford2022Whisper}. In the case of Polish recordings, transcription was performed with the automatic translation feature. Next, all transcriptions were manually verified by a bilingual native English speaker, and split between questions and answers. Full interviews transcriptions are available as a supplemental file.

\subsection{Data Processing}
\label{sec:DataProcessing}

Pre-processed interviews were placed in a spreadsheet format. Each of the categories for questions and answers was coded as a column. Aside from the main topics, interviewees also often provided additional information in a free-form manner - an attempt was made to categorize these responses as well:
\hypertarget{list:Categories}{}
\begin{enumerate*}[label=(\arabic*)]
    \item physical activity,
    \item psychology,
    \item science,
    \item coaching,
    \item in-game training,
    \item nutrition,
    \item sociology.
\end{enumerate*}

Next, four out of five authors as expert judges, separately grading the extracted abstract questions to alleviate possible biases towards contextual information. The judges were tasked with assigning a numerical value to each of the coupled question-answer contexts. Possible values were binary (either 0 - No, 1 - Yes, and additionally 2 - No Information) or Likert scale (1 - Strongly Disagree, 2 - Disagree, 3 - Neutral, 4 - Agree, 5 - Strongly Agree). The interview environment made it possible for the interviewer to ask multiple questions pertaining to the same topic. In the case when a similar question was asked after the initial question-answer context, such questions were judged separately and an average answer was calculated per interview.

\subsection{Post-processing}
\label{sec:post_processing}

After each of the referees provided their final graded sheet, further processing and numerical analyses were performed. Corresponding to the \hyperlink{list:Categories}{categories} presented in \nameref{sec:DataProcessing}, coding of the analyzed inteview questions is as follows: QA - physical activity, QP - psychology, QN - nutrition, QT - in-game training. Note that not all of the \hyperlink{list:Categories}{categories} could be represented numerically for the quantitative analysis.

\hypertarget{list:Questions}{}
\begin{itemize}[leftmargin=1.5cm]
    \item[\textbf{QA\_0:}]\label{item:QA_0} (likert) - Is physical activity important for players?
    \item[\textbf{QA\_1:}]\label{item:QA_1} (binary) - Do you train physically?
    \item[\textbf{QA\_2:}]\label{item:QA_2} (binary) - Do you work with a fitness coach?
    \item[\textbf{QP\_0:}]\label{item:QP_0} (likert) - Is psychology important for players?
    \item[\textbf{QP\_1:}]\label{item:QP_1} (binary) - Do you have a psychologist supporting you?
    \item[\textbf{QN\_0:}]\label{item:QN_0} (likert) - Is nutrition important for players?
    \item[\textbf{QN\_1:}]\label{item:QN_1} (binary) - Do you have a nutritionist supporting you?
    \item[\textbf{QN\_2:}]\label{item:QN_2} (binary) - Do you have a meal plan?
    \item[\textbf{QT\_0:}]\label{item:QT_0} (binary) - Do you make changes to your training before a tournament?
    \item[\textbf{QT\_1:}]\label{item:QT_1} (estimated time in hours) - How much time do you spend training in-game before a tournament?
    \item[\textbf{QS\_0:}]\label{item:QS_0} (likert) - How do you rate the atmosphere of the current tournament?
\end{itemize}
\section{Results}
\label{sec:Results}

\subsection{Quantitative Analysis}

\subsubsection{Physical Activity}

\begin{figure}[H]
    \centering
    \begin{subfigure}{.5\linewidth}
        \centering
        \includegraphics[width=\linewidth]{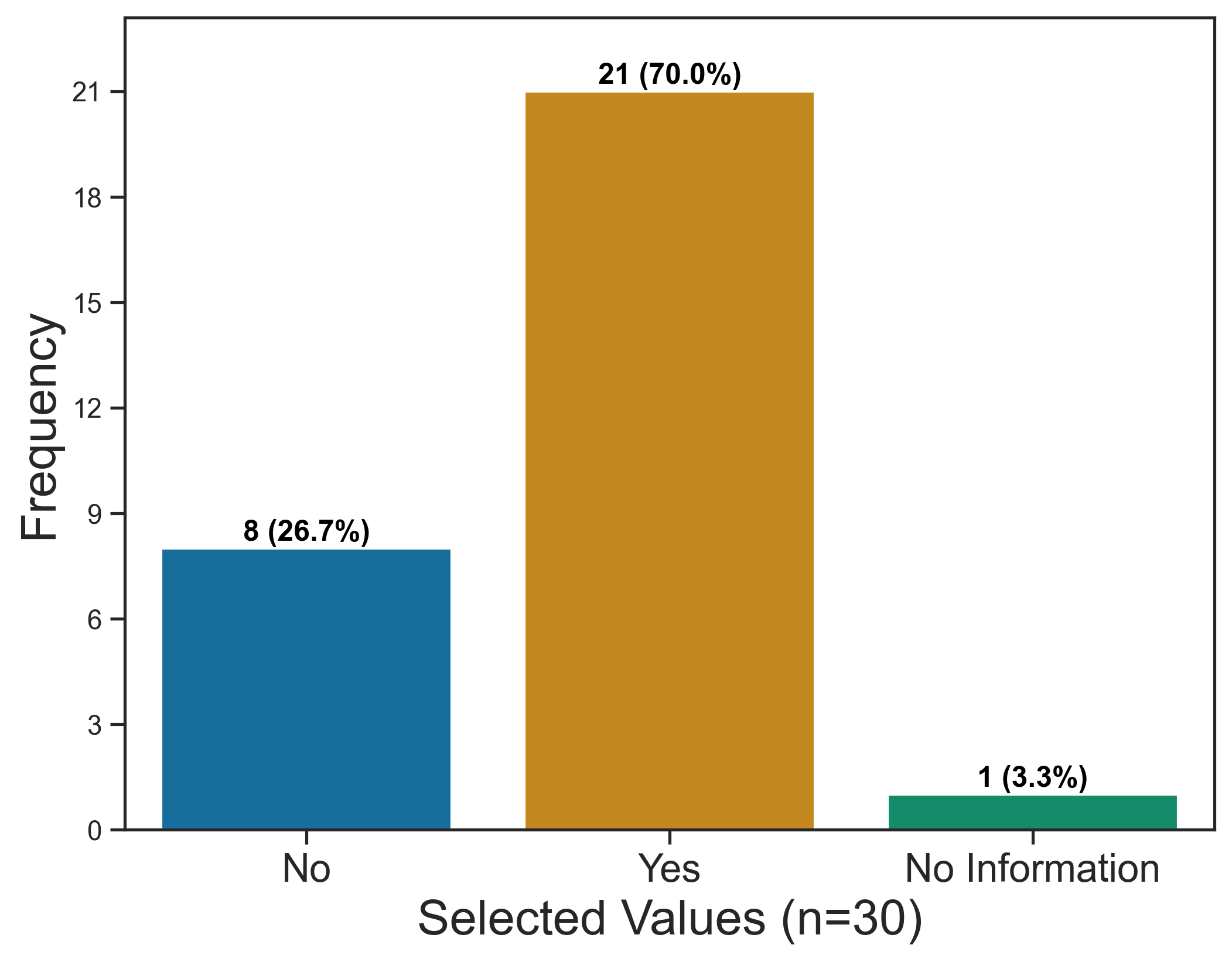}
        \caption{Choice frequency for ``Do you train physically?''.}
        \label{fig:sub1}
    \end{subfigure}\begin{subfigure}{.5\linewidth}
        \centering
        \includegraphics[width=\linewidth]{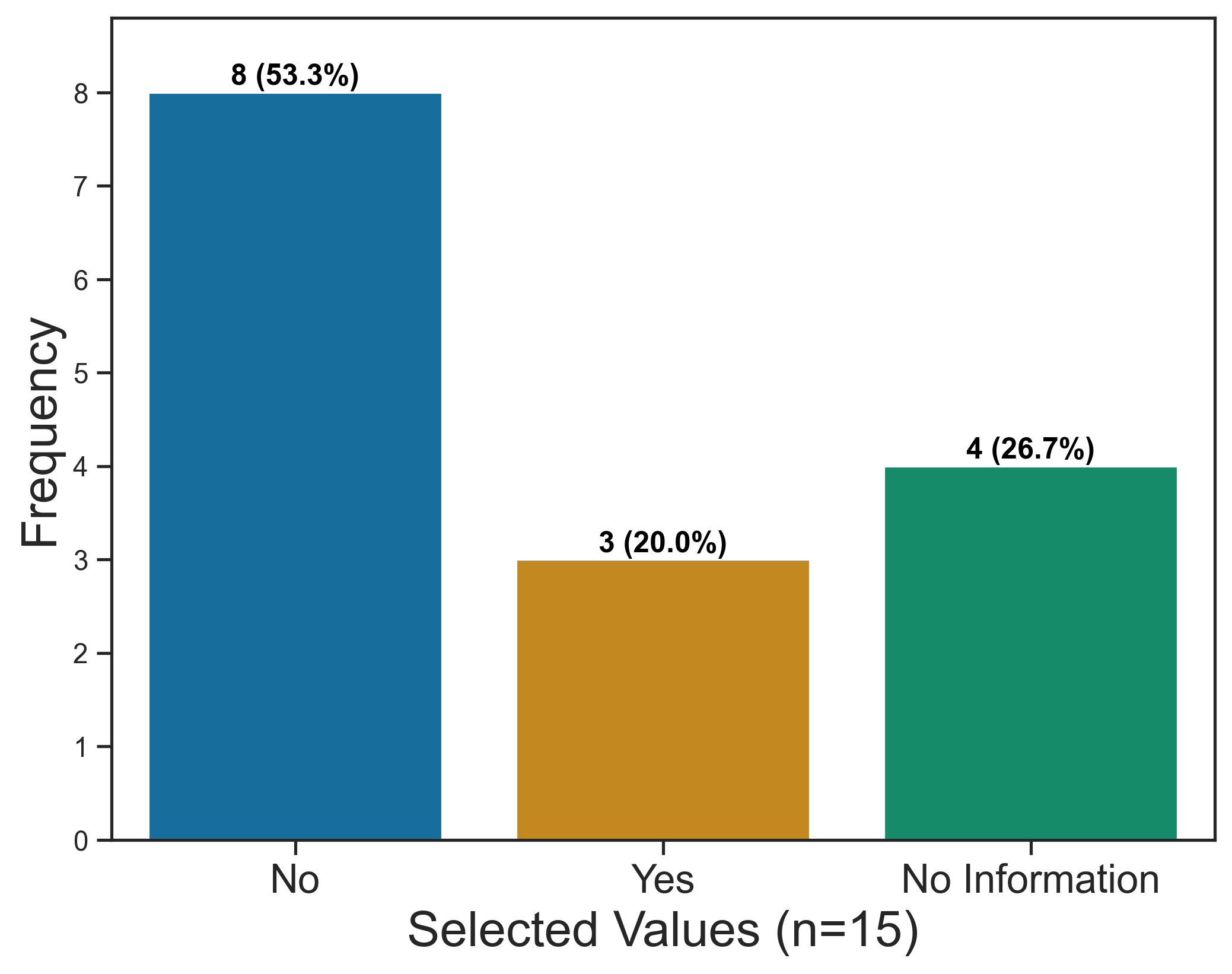}
        \caption{Choice frequency for ``Do you work with a fitness coach?''.}
        \label{fig:sub2}
    \end{subfigure}\vspace{1em} \begin{subfigure}{\linewidth}
        \centering
        \includegraphics[width=0.8\linewidth]{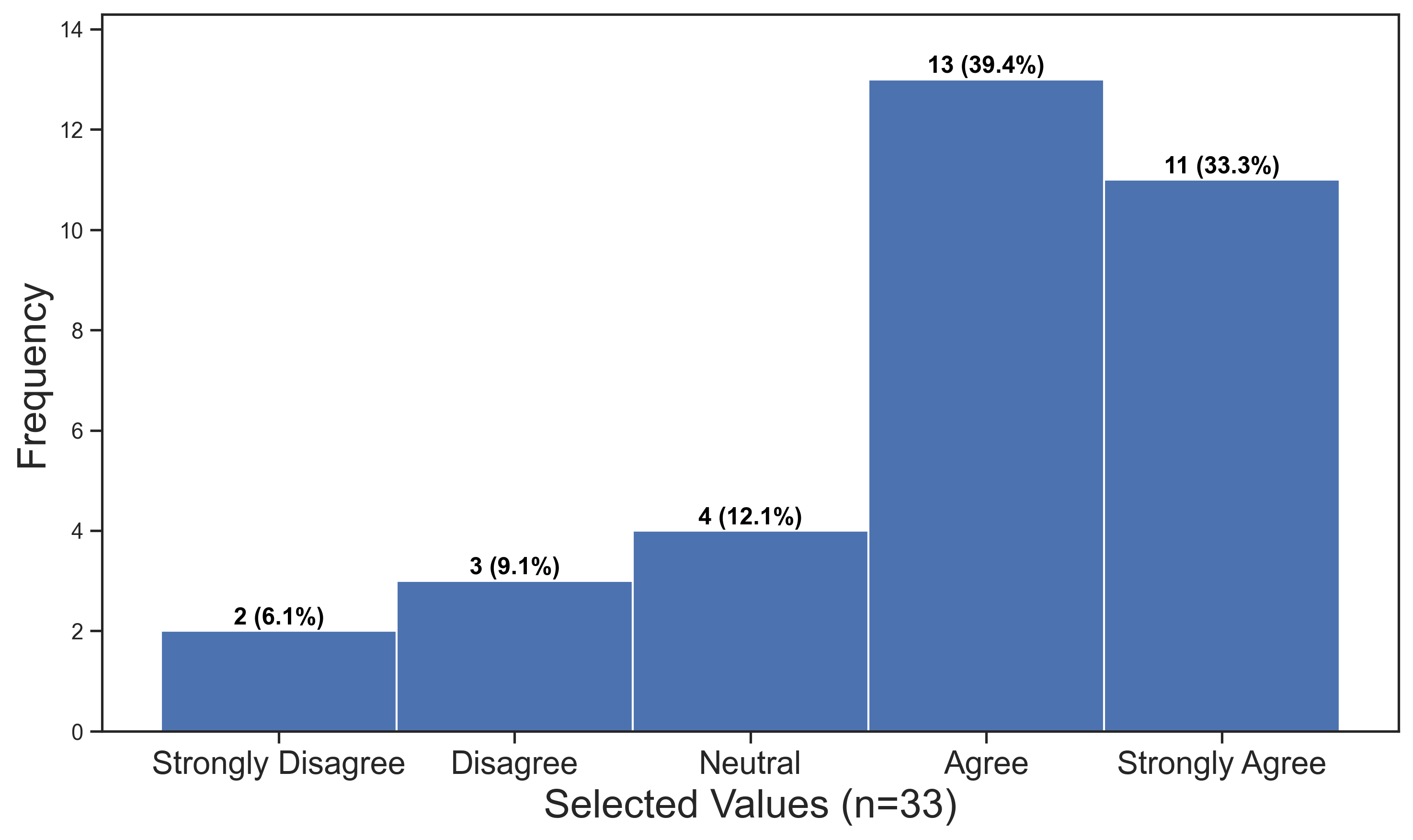}
        \caption{Choice frequency for ``Is physical activity important for players?''.}
        \label{fig:sub3}
    \end{subfigure}\caption{Visualization for all coded questions related to physical activity.}
    \label{fig:ActivityPlots}
\end{figure}

Aggregated referee responses are visualized in \autoref{fig:ActivityPlots}. Visual investigation of the plots indicates that when asked about their approach towards physical activity - \autoref{fig:sub3}, the majority of interviewees stated that they Agree or Strongly Agree that it can positively influence their gameplay. Additionally, most of the players - \autoref{fig:sub1}, indicated that they train physically. On the other hand - \autoref{fig:sub2}, when asked about the external support from a professional fitness coach, they typically indicated not having such support at the time.

\subsubsection{Psychology}

\begin{figure}[H]
    \centering
    \begin{subfigure}{.5\linewidth}
        \centering
        \includegraphics[width=\linewidth]{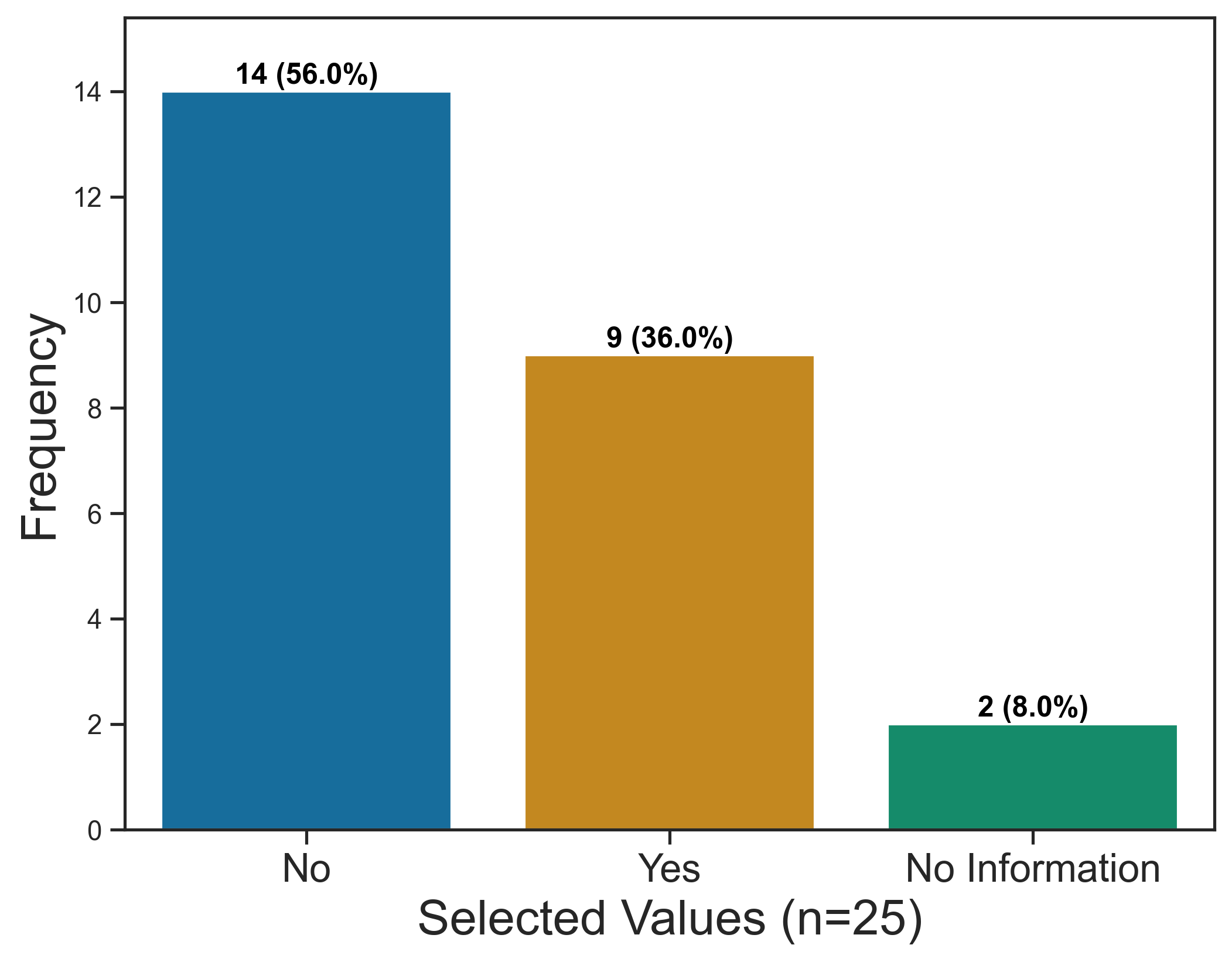}
        \caption{Choice frequency for ``Do you have a psychologist supporting you?''.}
        \label{fig:psych_sub1}
    \end{subfigure}\vspace{1em} \begin{subfigure}{\linewidth}
        \centering
        \includegraphics[width=0.8\linewidth]{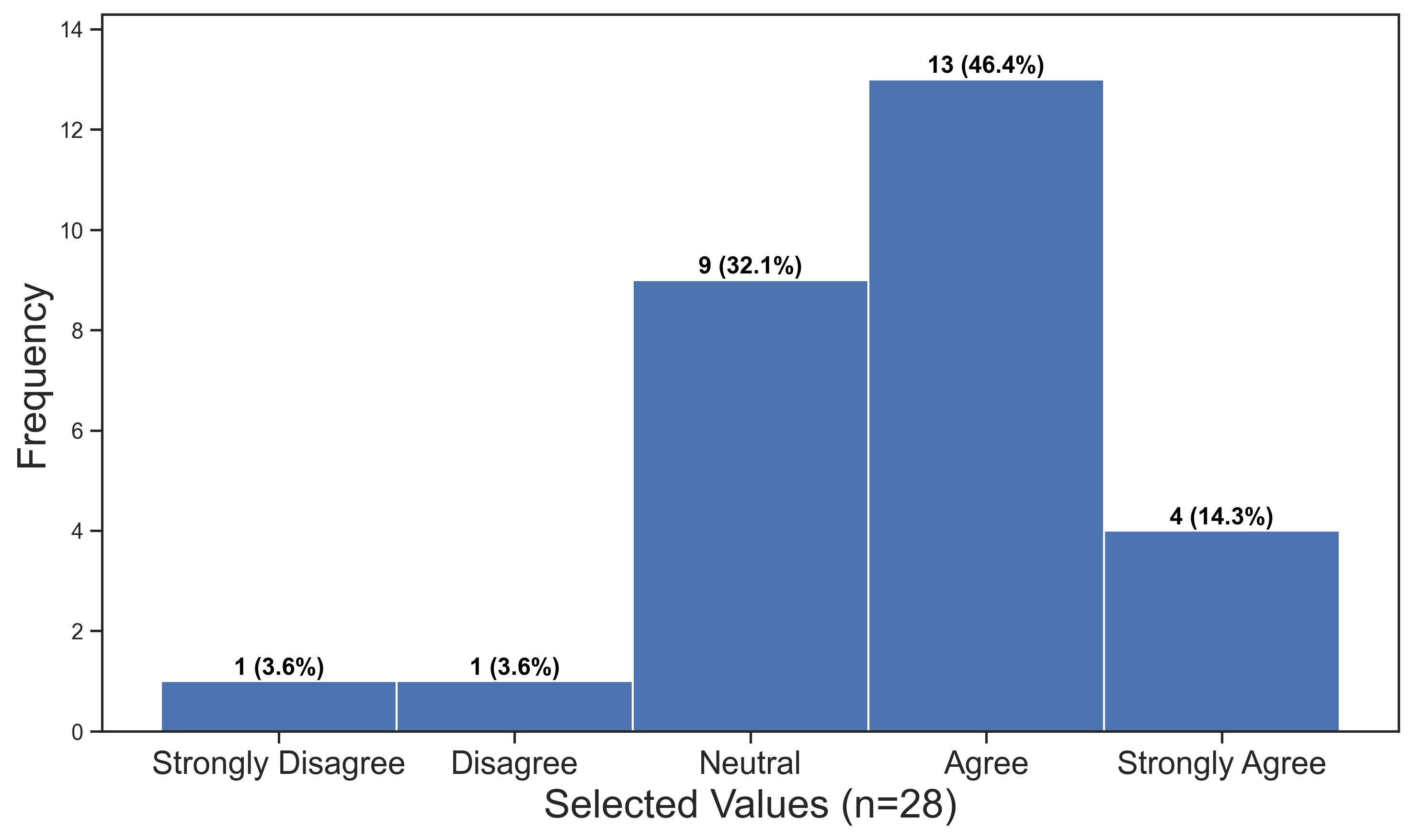}
        \caption{Choice frequency for ``Is psychology important for players?''.}
        \label{fig:psych_sub2}
    \end{subfigure}\caption{Visualization for all coded questions related to psychology.}
    \label{fig:PsychologyPlots}
\end{figure}

When asked about psychological aspects - \autoref{fig:PsychologyPlots}, the majority of the interviewees agreed that it can be important for success in esports \autoref{fig:psych_sub2}. On the other hand, when asked if they have access to a professional psychological support, most players responded negatively - \autoref{fig:psych_sub1}.

\subsubsection{Nutrition}

\begin{figure}[H]
    \centering
    \begin{subfigure}{.5\linewidth}
        \centering
        \includegraphics[width=\linewidth]{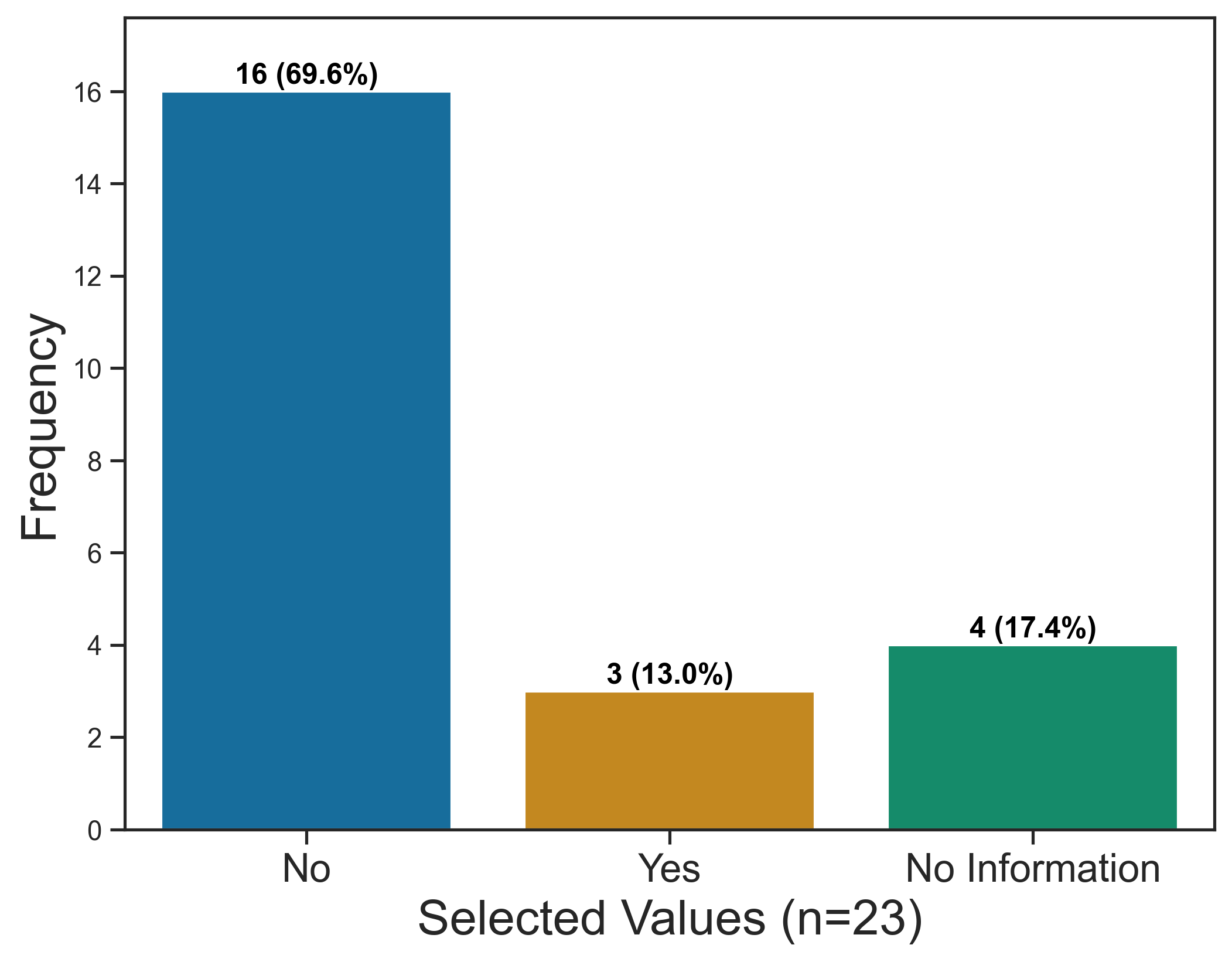}
        \caption{Choice frequency for ``Do you have a nutritionist supporting you?''.}
        \label{fig:nutrition_sub1}
    \end{subfigure}\begin{subfigure}{.5\linewidth}
        \centering
        \includegraphics[width=\linewidth]{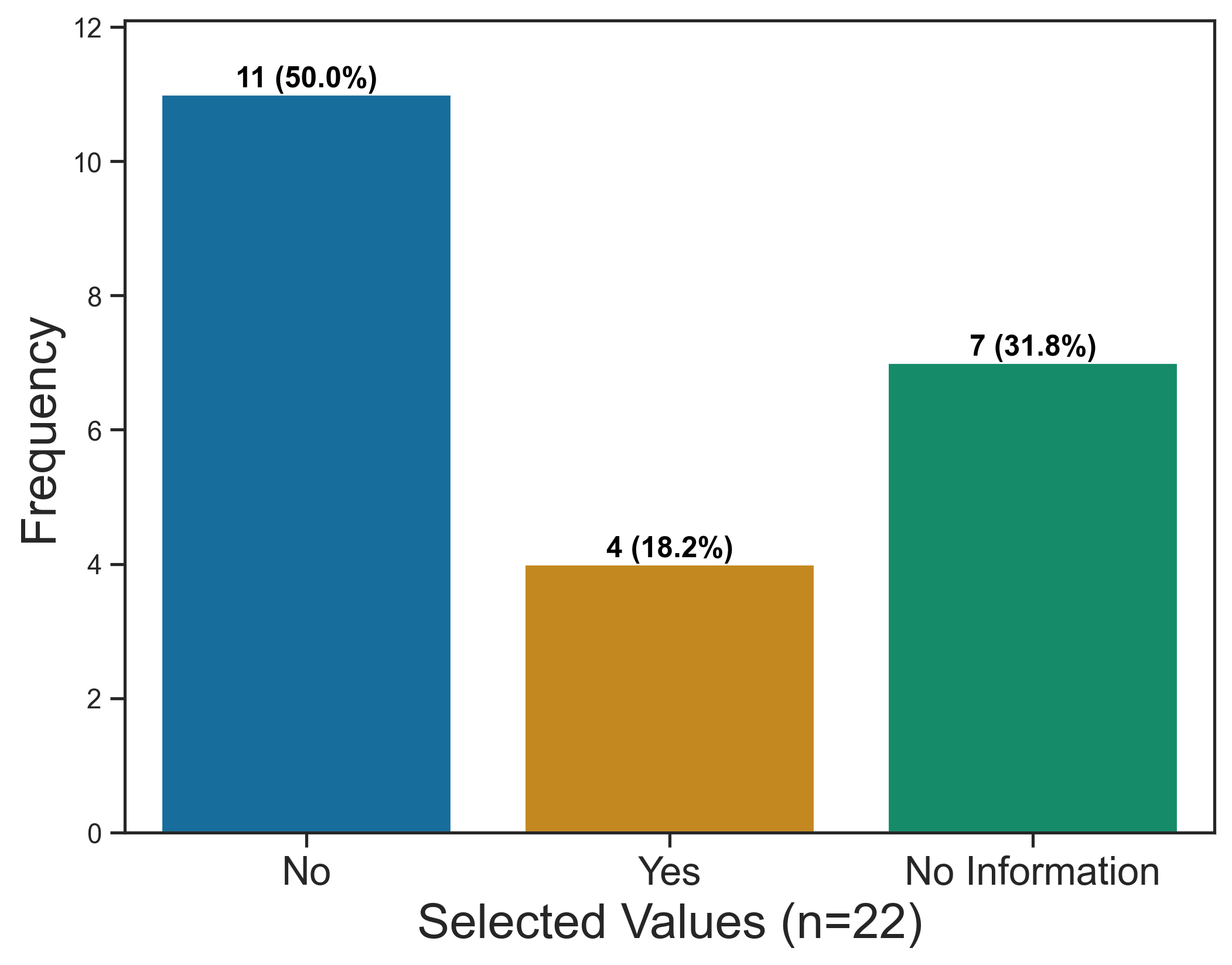}
        \caption{Choice frequency for ``Do you have a meal plan?''.}
        \label{fig:nutrition_sub2}
    \end{subfigure}\vspace{1em} \begin{subfigure}{\linewidth}
        \centering
        \includegraphics[width=0.8\linewidth]{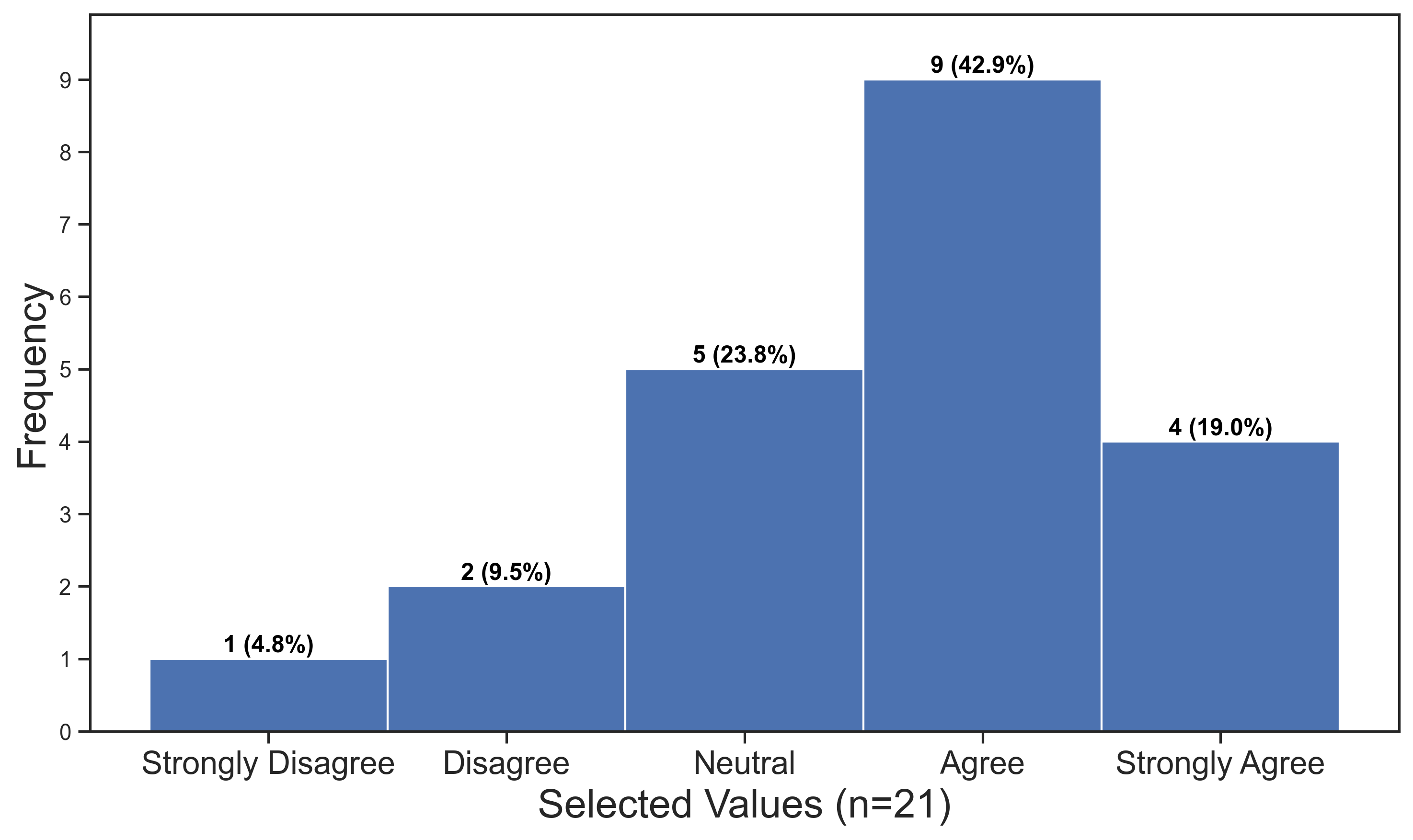}
        \caption{Choice frequency for ``Is nutrition important for players?''.}
        \label{fig:nutrition_sub3}
    \end{subfigure}\caption{Visualization for all coded questions related to nutrition.}
    \label{fig:NutritionPlots}
\end{figure}

While discussing nutrition and its possible influence on esports performance - \autoref{fig:NutritionPlots}, players indicated that they either Agree or Strongly Agree with the premise of nutrition being important for their performance - \autoref{fig:nutrition_sub3}. Despite these statements, most of the players indicated that they do not follow any meal plan and do not consult with a professional nutritionist, as can be seen in \autoref{fig:nutrition_sub1} and \autoref{fig:nutrition_sub2}

\subsubsection{Response Comparison}

To compare the perceived responses between different coded question categories, we have conducted a Mann-Whitney U test for all of the likert scale questions. The results of the test are presented in \autoref{tab:question_comparison}.

\begin{table}[H]
    \caption{Results of Mann-Whitney U test for likert scale coded questions.}
    \label{tab:question_comparison}
    \centering
    \begin{tabular}{lllllll}
        \hline
        \multicolumn{1}{c}{\multirow{2}{*}{question}} & \multicolumn{2}{c}{QA\_0} & \multicolumn{2}{c}{QP\_0} & \multicolumn{2}{c}{QN\_0}                                              \\ \cline{2-7}
        \multicolumn{1}{c}{}                          & u-stat                    & p value                   & u-stat                    & p-value & u-stat                 & p-value \\ \hline
        QA\_0                                         & \multicolumn{2}{c}{--}    &                           &                           &         &                                  \\
        QP\_0                                         & 379.0                     & 0.207                     & \multicolumn{2}{c}{--}    &         &                                  \\
        QN\_0                                         & 292.5                     & 0.318                     & 297.5                     & 0.948   & \multicolumn{2}{c}{--}           \\ \hline
    \end{tabular}
\end{table}

Based on these results we conclude that all of the questions were answered similarly indicating that esports professionals have positive attitude towards the statement that either physical activity, psychology, or nutrition is important to their performance.

\subsubsection{In-Game Training}

\begin{figure}[H]
    \centering
    \begin{subfigure}{.5\linewidth}
        \centering
        \includegraphics[width=\linewidth]{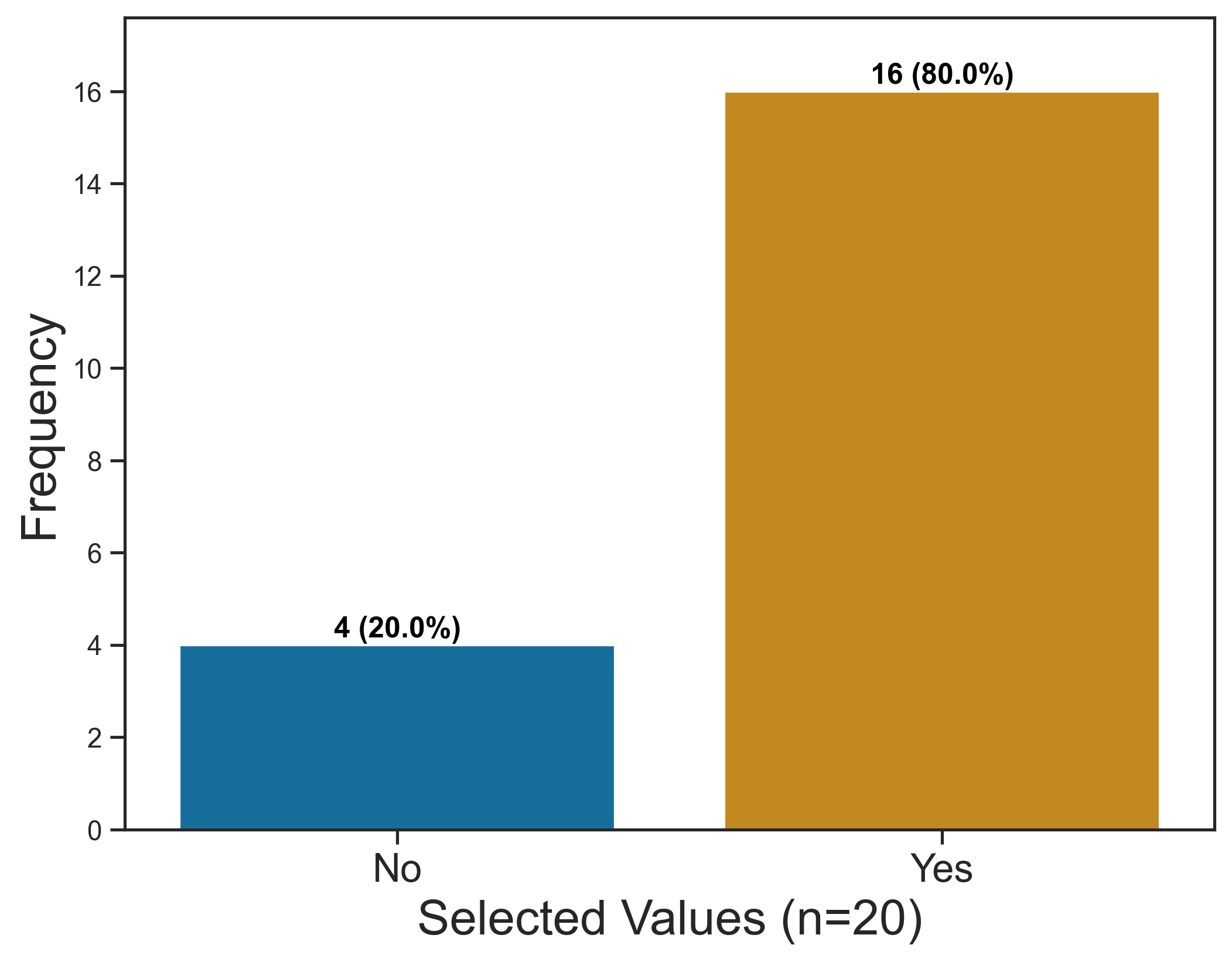}
        \caption{Choice frequency for ``Do you make changes to your training before a tournament?''.}
        \label{fig:training_sub1}
    \end{subfigure}\vspace{1em} \begin{subfigure}{\linewidth}
        \centering
        \includegraphics[width=0.8\linewidth]{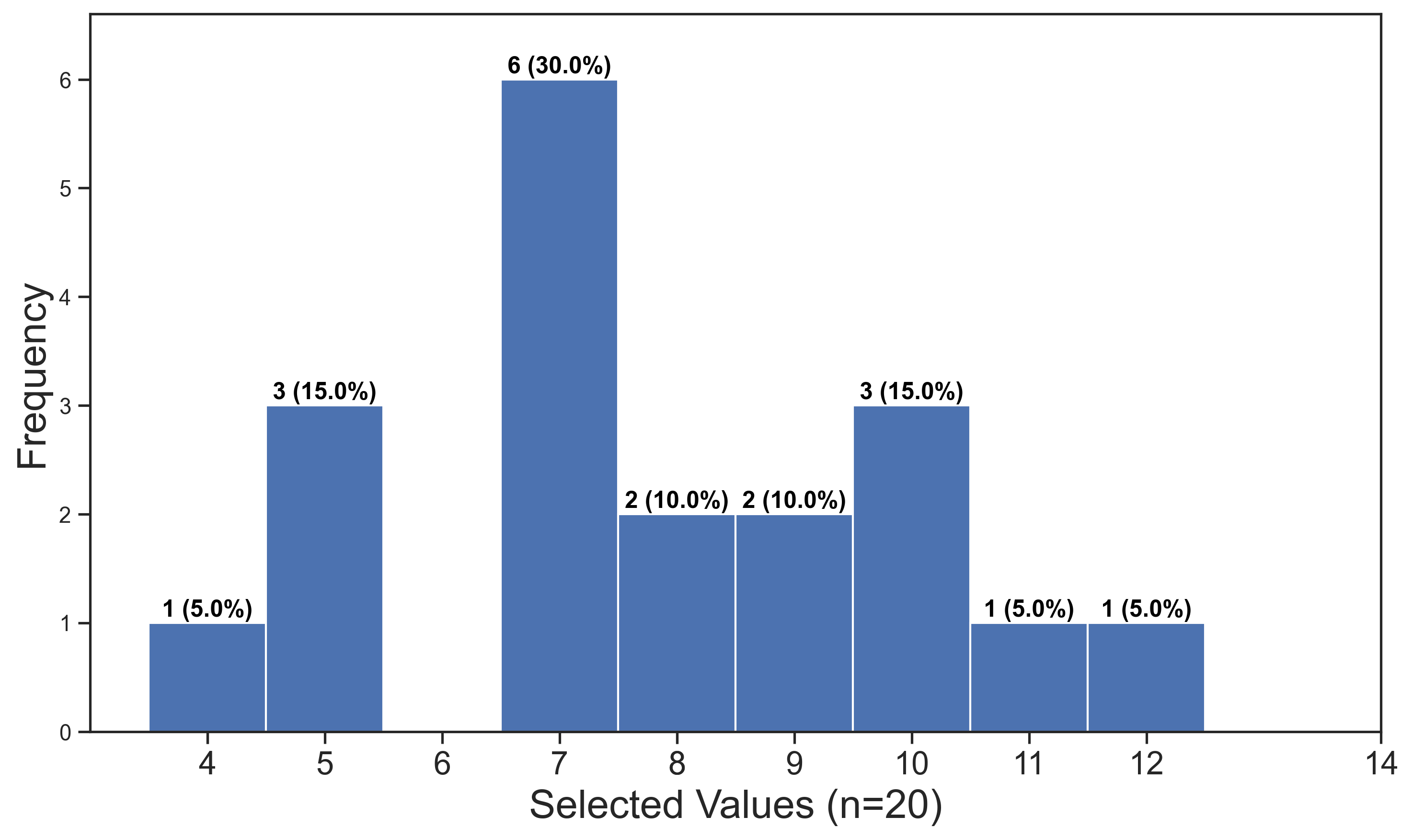}
        \caption{Choice frequency for ``How much time do you spend training in-game before a tournament?''.}
        \label{fig:training_sub2}
    \end{subfigure}\caption{Visualization for all coded questions related to psychology.}
    \label{fig:TrainingPlots}
\end{figure}

When asked about the in-game training characteristis, interviewees indicated that they do change how they train when in preparation for a tournament - \autoref{fig:TrainingPlots}. Majority of the respondents indicated average training time between 7 and 9 hours being most prevalent.

\subsection{Tournament Atmosphere}

When asked about the tournament atmosphere (n=18), i.e. ``How do you rate the atmosphere of the current tournament?'', the majority of the players indicated that they strongly agree with the statement (n=8), others (n=6) stated that they agree, and finally (n=4) provided neutral statements.

\subsection{Qualitative Analysis}

As discussed in \autoref{sec:post_processing}, not all questions could be represented numerically, and yet key information was provided in their context.

\subsubsection{Recommendations for Aspiring Players}

When asked about what are their recommendations for the younger players that would like to try and become professional, interviewee 16 stated:
\begin{quote}
    Like I said before, just practice, practice, practice, that's all it really comes down to and just having a good mindset. Those are the two main things I'd definitely say, yeah.
\end{quote}

In this statement the player underlined the idea of practice and having the right mindset. Interviewee~17 stated:
\begin{quote}
    I'd say I was doing the basics. I was not doing more than I could have done. I was just working enough to graduate, not to do more. So I'd say it's up to everyone, every each person to study as much as they want or so. Maybe some people are not going to school to play. I was doing that sometimes but that's not what you should do obviously. It's up to everyone to work as much as you want and then to try to play outside from the school hours.
\end{quote}

Which emphasized that aspiring players should play as much as possible outside of other responsibilities. This is in line with what interviewee~20 stated:
\begin{quote}
    You know what, it depends on who is at home (the environment). When parents are always going to put pressure on formal education, and they won't let their son (child) fullfill themselves in e-gaming (gaming, esports), it will be a very difficult path, but you have to go through it and be persistent, persistent, persistent. Do not give up. After two or three years, if you want to become a professional player, sooner or later you can make it but it is very hard.
\end{quote}

Which underlined the importance of environmental factors as well as being persistent as key indicators that play a role in esports success. Similarly, interviewee~29 stated:
\begin{quote}
    It's just keep working, you should just finish school, you can play when you get home and with your team in the evening, it's more like the social life you sacrifice because you have work, you have school, some things you just have to do if you don't succeed and that's just the fact and just keep trying and go forward but don't go all in if you know what I mean ...
\end{quote}

Which additionally mentions not to sacrifice everything for the sake of an esports career. On top of that, interviewee~31 stated:
\begin{quote}
    I feel like you have to really work hard. I think there's a lot of people in the space that are really, really good at video games, but to get to a professional level, I think the biggest thing is stopping people is themselves, and you can be really good, but to become a professional is you need to be able to take criticisms, be able to send maybe your match demo to somebody and be like, hey, where did I mess up, and be willing to listen to that. You're not like, oh, this guy's an idiot, he doesn't understand, or whatever, you have to be able to take criticisms from your peers, from the fans, from your teammates, and I think that's kind of the difference between a professional and a really good player is how serious they're willing to take themselves and how much time they're willing to put into it. So in terms of tools, I think there is definitely tools out there that can help players grow, but in my personal opinion, your biggest enemy is yourself.
\end{quote}

Putting emphasis on the psychological and communication factors playing a role in the success in esports. Furthermore, interviewee~39 stated:
\begin{quote}
    There is no secret if you're good, you're gonna get seen and and you need you know a bit of luck. Plus being very good. So It is just a matter of, Yeah, making things happen just show yourself as much as possible. And if you're very good, you're gonna get seen and you're gonna climb up the ladder in a way and and yeah work hard
\end{quote}

All of the interviewees providing recommendations for aspiring esports players indicated hard work, persistence, or environmental factors coupled with luck as key indicators for success in esports.

\subsubsection{Tools and Esports Science}

When asked about the availability of tools, and current state of educational and scientifically applicable information, interviewee~12 stated:
\begin{quote}
    Difficult question. I've never looked for any advice (tutorials) on how to train in esport. I've watched some videos, as Piotr said, from Virtus.pro, how they prepare, how they train. Of course, it's about the Counter-Strike team, but the influence of science... Personally I've not taken any scientific advice from articles, bachelor, or some master thesis on this topic.
\end{quote}
And later provided more context with:
\begin{quote}
    Yes, for sure there are some statistics about how an average player from top teams spens their time in training. Maybe even on how they eat, or if they do something besides that. So surely some statistics could show that something out there works and helps in these preparations (training). But as I said, I've never looked for any statistics, I've never supported myself in this way, or looked for any advice.
\end{quote}
At the same time indicating that the statistics are important in discovering weaknesses in gameplay:
\begin{quote}
    Yes, we do. We take care of statistics, mainly percentage of wins of teams from different leagues, from different maps, and on different sides of the map. But I don't think it changes... If we are weak in the map, we train it or we think that we don't want to play this map and we always ban it. But specific preparation outside of the game and on a specific map, we don't do that. There is no such thing that we have to focus more on the days when we play with these opponents. There isn't anything like that.
\end{quote}

On the other hand when a performacne coach (interviewee~15) of one of the teams was asked about the availability of the tools, these were his statements:
\begin{quote}
    I mean, mainly there are websites that collect the data and statistics, there are some software tools that are paid, and unfortunately they cost a lot. We try to use them. I think that there is no dedicated tool, you just got to know what you want to verify, what works and what doesn't. Additionally, the tools for physical training are mostly available online. If it is tools for stretching, the balls for massaging the muscles? And in the game, well, it's a computer. I do not see any other essential tools.
\end{quote} \section{Limitations}
\label{sec:Limitations}

It should be noted that the interviews were conducted in a short form, in a fast paced tournament environment. Therefore, there was no way of coduncting a full in-depth analysis with existing sociological methodologies. Moreover, as mentioned in the \nameref{sec:interview_collection} section, the data was collected from 2016 through 2019 and in the everchanging esports landscape these represent only a very limited sample of top tournament players, coaches, and managers. Despite that, we think that our work offers a unique snapshot of the state of top-level esports at the time of data gathering. Additionally, as opinions may of course change over time, we cannot verify if the interviewees still hold the same stance on the investigated topics.

\section{Summary}
\label{sec:summary}

Despite some of the players indicating that physical activity is not relevant in esports, there are results that contradict such anecdotal experience. Breaking prolonged sitting with a six minute walk has shown to improve the executive functioning of players \cite{DiFrancisco2021}. Further, the widespread consensus on physical activity is that it is a positive stimulus for human cognitive performance \cite{Mandolesi2018,Brown2021}. Yet it is unclear what the optimal training load and intensity are to facilitate positive effects on in-game performance. This includes potential scheduling issues, as e.g. intense physical training just prior to high levels of mental activity may even be detrimental to performance.

Spitzer et al. \cite{Spitzer2022} suggest a theoretical aproach for the possibility of knowledge transfer by AI systems. Learning in new environments may include acquiring "tacit" knowledge, which is hard to express by available language but could be captured in data. On the other hand, as the esport research field progresses this knowledge should be documented, verified, and applied to future esport players and conventional sports.

\subsection{Future Research}

We recommend further research into the broad area of esports to answer the following questions:
\begin{enumerate*}[label=(\arabic*)]
    \item What should be the load and intensity of physical activity training to significantly boost the performance of esports players?
    \item How to structure esports training so that it can fit physical activity in a non-invasive way?
    \item What should be the methodical structure of training and practice in esports to maximize human performance?
    \item What are the currently available training tools to manage training load and intensity?
    \item Is it possible to use online and offline AI systems in esports training effectively?
\end{enumerate*}
\section{Conclusions}
\label{sec:Conclusions}

Given our results from the quantitative analyses, we have concluded that most of the players and esports staff were aware of the potential positive effects of physical activity, psychology, and nutrition (\textbf{RQ 1}). Additionally, a higher percentage of interviewees indicated that they strongly agree with physical activity being important, than for the nutrition or psychology. Despite the high-profile of the interviewees, it is clear the they had limited access to support staff to take care of these aspects for them (\textbf{RQ 2}). On the other hand, 36\% of the respondents indicated that they have access to a psychologist which compared other questions: fitness coach - 20\%, or a nutritionist - 13\%. Juxtaposing these claims with the 70\% of interviewees stating that they train physically, and only 18.2\% following a meal plan we conclude that physical activity and psychological aspects are of key importance to players.

Based on the players recommendations (\textbf{RQ 3}), we conclude that the best way for the aspiring players to reach their goals in esports is to be persistent, hard working, and not sacrifice other areas of life if possible. This points towards a holistic view of esports represented by the top-level athletes. Additionally, respondents indicated that they use statistical software to track gameplay weaknesses, and look for various educational materials on how to play better.
\section*{Acknowledgements}

We would like to acknowledge various contributions by the members of the community, namely: Oliwia Kamińska, Konrad Daroch, Kacper Zyzak, and others that helped in the interview collection process.

\subsection*{Authors' contributions}

\begin{itemize}
    \item Conceptualization: Andrzej Białecki;
    \item Supervision: Andrzej Białecki, Jan Gajewski;
    \item Methodology: Andrzej Białecki, Paweł Dobrowolski;
    \item Investigation: Andrzej Białecki, Peter Xenopoulos, Paweł Dobrowolski, Robert Białecki, Jan Gajewski;
    \item Writing - original draft: Andrzej Białecki, Peter Xenopoulos;
    \item Writing - review and editing: Andrzej Białecki, Paweł Dobrowolski, Robert Białecki;
    \item Technical Oversight: Andrzej Białecki;
    \item Experiments: Andrzej Białecki;
\end{itemize}

\section*{Declarations}

Authors declare no conflict of interest.

\bibliographystyle{IEEEtran}
\bibliography{sources.bib}

\begin{thebibliography}{10}
\providecommand{\url}[1]{#1}
\csname url@samestyle\endcsname
\providecommand{\newblock}{\relax}
\providecommand{\bibinfo}[2]{#2}
\providecommand{\BIBentrySTDinterwordspacing}{\spaceskip=0pt\relax}
\providecommand{\BIBentryALTinterwordstretchfactor}{4}
\providecommand{\BIBentryALTinterwordspacing}{\spaceskip=\fontdimen2\font plus
\BIBentryALTinterwordstretchfactor\fontdimen3\font minus
  \fontdimen4\font\relax}
\providecommand{\BIBforeignlanguage}[2]{{%
\expandafter\ifx\csname l@#1\endcsname\relax
\typeout{** WARNING: IEEEtran.bst: No hyphenation pattern has been}%
\typeout{** loaded for the language `#1'. Using the pattern for}%
\typeout{** the default language instead.}%
\else
\language=\csname l@#1\endcsname
\fi
#2}}
\providecommand{\BIBdecl}{\relax}
\BIBdecl

\bibitem{BialeckiRedefiningSports2022}
\BIBentryALTinterwordspacing
A.~Białecki, R.~Białecki, and J.~Gajewski, ``{Redefining Sports: Esports,
  Environments, Signals and Functions},'' \emph{International Journal of
  Electronics and Telecommunications}, vol. vol. 68, no. No 3, pp. 541--548,
  aug 2022. [Online]. Available: \url{http://doi.org/10.24425/ijet.2022.141272}
\BIBentrySTDinterwordspacing

\bibitem{Fried2023}
\BIBentryALTinterwordspacing
G.~Fried, ``Esports minus sport?'' \emph{Journal of Electronic Gaming and
  Esports}, vol.~1, no.~1, pp. jege.2023--0018, 2023. [Online]. Available:
  \url{https://doi.org/10.1123/jege.2023-0018}
\BIBentrySTDinterwordspacing

\bibitem{Freeman2017}
\BIBentryALTinterwordspacing
G.~Freeman and D.~Y. Wohn, ``Esports as an emerging research context at chi:
  Diverse perspectives on definitions,'' in \emph{Proceedings of the 2017 CHI
  Conference Extended Abstracts on Human Factors in Computing Systems}, ser.
  CHI EA '17.\hskip 1em plus 0.5em minus 0.4em\relax New York, NY, USA:
  Association for Computing Machinery, 2017, pp. 1601--1608. [Online].
  Available: \url{https://doi.org/10.1145/3027063.3053158}
\BIBentrySTDinterwordspacing

\bibitem{URLPei2019}
\BIBentryALTinterwordspacing
A.~Pei, ``This esports giant draws in more viewers than the super bowl, and
  it's expected to get even bigger,'' Apr 2019, accessed: 2024.09.10. [Online].
  Available:
  \url{https://www.cnbc.com/2019/04/14/league-of-legends-gets-more-viewers-than-super-bowlwhats-coming-next.html}
\BIBentrySTDinterwordspacing

\bibitem{URLEsportsCharts}
``{E}sports {C}harts - {E}sports {V}iewership, {P}opularity and {A}nalytics ---
  escharts.com,'' \url{https://escharts.com/}, accessed: 2024.09.10.

\bibitem{Holden2017}
\BIBentryALTinterwordspacing
J.~T. Holden, A.~Kaburakis, and R.~Rodenberg, ``The future is now: Esports
  policy considerations and potential litigation,'' \emph{Journal of Legal
  Aspects of Sport}, vol.~27, no.~1, pp. 46--78, Feb. 2017. [Online].
  Available: \url{https://doi.org/10.1123/jlas.2016-0018}
\BIBentrySTDinterwordspacing

\bibitem{Rudolf2020}
\BIBentryALTinterwordspacing
K.~Rudolf, P.~Bickmann, I.~Froböse, C.~Tholl, K.~Wechsler, and C.~Grieben,
  ``Demographics and health behavior of video game and esports players in
  germany: The esports study 2019,'' \emph{International Journal of
  Environmental Research and Public Health}, vol.~17, no.~6, 2020. [Online].
  Available: \url{https://doi.org/10.3390/ijerph17061870}
\BIBentrySTDinterwordspacing

\bibitem{Arnau2023}
\BIBentryALTinterwordspacing
A.~Baena-Riera, L.~M. Carrani, A.~Piedra, and J.~Peña, ``Exercise
  recommendations for e-athletes: Guidelines to prevent injuries and health
  issues,'' \emph{Journal of Electronic Gaming and Esports}, vol.~1, no.~1, pp.
  jege.2023--0003, 2023. [Online]. Available:
  \url{https://doi.org/10.1123/jege.2023-0003}
\BIBentrySTDinterwordspacing

\bibitem{Madden2021}
\BIBentryALTinterwordspacing
D.~Madden and C.~Harteveld, ``“constant pressure of having to perform”:
  Exploring player health concerns in esports,'' in \emph{Proceedings of the
  2021 CHI Conference on Human Factors in Computing Systems}, ser. CHI
  '21.\hskip 1em plus 0.5em minus 0.4em\relax New York, NY, USA: Association
  for Computing Machinery, 2021. [Online]. Available:
  \url{https://doi.org/10.1145/3411764.3445733}
\BIBentrySTDinterwordspacing

\bibitem{Hamilton2014}
\BIBentryALTinterwordspacing
W.~A. Hamilton, O.~Garretson, and A.~Kerne, ``Streaming on twitch: Fostering
  participatory communities of play within live mixed media,'' in
  \emph{Proceedings of the SIGCHI Conference on Human Factors in Computing
  Systems}, ser. CHI '14.\hskip 1em plus 0.5em minus 0.4em\relax New York, NY,
  USA: Association for Computing Machinery, 2014, pp. 1315--1324. [Online].
  Available: \url{https://doi.org/10.1145/2556288.2557048}
\BIBentrySTDinterwordspacing

\bibitem{Torres2022}
\BIBentryALTinterwordspacing
A.~Torres-Toukoumidis, \emph{Esports and the Media}.\hskip 1em plus 0.5em minus
  0.4em\relax Routledge, jul 2022. [Online]. Available:
  \url{https://doi.org/10.4324/9781003273691}
\BIBentrySTDinterwordspacing

\bibitem{Kow2013}
\BIBentryALTinterwordspacing
Y.~M. Kow and T.~Young, ``Media technologies and learning in the starcraft
  esport community,'' in \emph{Proceedings of the 2013 Conference on Computer
  Supported Cooperative Work}, ser. CSCW '13.\hskip 1em plus 0.5em minus
  0.4em\relax New York, NY, USA: Association for Computing Machinery, 2013, pp.
  387--398. [Online]. Available: \url{https://doi.org/10.1145/2441776.2441821}
\BIBentrySTDinterwordspacing

\bibitem{Watson2021}
\BIBentryALTinterwordspacing
B.~Watson, J.~Spjut, J.~Kim, J.~Listman, S.~Kim, R.~Wimmer, D.~Putrino, and
  B.~Lee, ``Esports and high performance hci,'' in \emph{Extended Abstracts of
  the 2021 CHI Conference on Human Factors in Computing Systems}, ser. CHI EA
  '21.\hskip 1em plus 0.5em minus 0.4em\relax New York, NY, USA: Association
  for Computing Machinery, 2021. [Online]. Available:
  \url{https://doi.org/10.1145/3411763.3441313}
\BIBentrySTDinterwordspacing

\bibitem{Chiu2021}
\BIBentryALTinterwordspacing
W.~Chiu, T.~C.~M. Fan, S.-B. Nam, and P.-H. Sun, ``{Knowledge Mapping and
  Sustainable Development of eSports Research: A Bibliometric and Visualized
  Analysis},'' \emph{Sustainability}, vol.~13, no.~18, 2021. [Online].
  Available: \url{https://doi.org/10.3390/su131810354}
\BIBentrySTDinterwordspacing

\bibitem{Tabacof2021}
\BIBentryALTinterwordspacing
L.~Tabacof, S.~Dewil, J.~E. Herrera, M.~Cortes, and D.~Putrino, ``Adaptive
  esports for people with spinal cord injury: New frontiers for inclusion in
  mainstream sports performance,'' \emph{Frontiers in Psychology}, vol.~12,
  2021. [Online]. Available: \url{https://doi.org/10.3389/fpsyg.2021.612350}
\BIBentrySTDinterwordspacing

\bibitem{Sheldon2003}
\BIBentryALTinterwordspacing
N.~Sheldon, E.~Girard, S.~Borg, M.~Claypool, and E.~Agu, ``The effect of
  latency on user performance in warcraft iii,'' ser. NetGames '03.\hskip 1em
  plus 0.5em minus 0.4em\relax New York, NY, USA: Association for Computing
  Machinery, 2003, pp. 3--14. [Online]. Available:
  \url{https://doi.org/10.1145/963900.963901}
\BIBentrySTDinterwordspacing

\bibitem{Kriglstein2021}
\BIBentryALTinterwordspacing
S.~Kriglstein, A.~L. Martin-Niedecken, L.~Turmo~Vidal, M.~Klarkowski,
  K.~Rogers, S.~Turkay, M.~Seif El-Nasr, E.~M\'{a}rquez~Segura, A.~Drachen, and
  P.~H\"{a}m\"{a}l\"{a}inen, ``Special interest group: The present and future
  of esports in hci,'' in \emph{Extended Abstracts of the 2021 CHI Conference
  on Human Factors in Computing Systems}, ser. CHI EA '21.\hskip 1em plus 0.5em
  minus 0.4em\relax New York, NY, USA: Association for Computing Machinery,
  2021. [Online]. Available: \url{https://doi.org/10.1145/3411763.3450402}
\BIBentrySTDinterwordspacing

\bibitem{Wu2021}
\BIBentryALTinterwordspacing
M.~Wu, J.~S. Lee, and C.~Steinkuehler, ``Understanding tilt in esports: A study
  on young league of legends players,'' in \emph{Proceedings of the 2021 CHI
  Conference on Human Factors in Computing Systems}, ser. CHI '21.\hskip 1em
  plus 0.5em minus 0.4em\relax New York, NY, USA: Association for Computing
  Machinery, 2021. [Online]. Available:
  \url{https://doi.org/10.1145/3411764.3445143}
\BIBentrySTDinterwordspacing

\bibitem{Selen2020}
\BIBentryALTinterwordspacing
S.~T\"{u}rkay, J.~Formosa, S.~Adinolf, R.~Cuthbert, and R.~Altizer, ``See no
  evil, hear no evil, speak no evil: How collegiate players define, experience
  and cope with toxicity,'' in \emph{Proceedings of the 2020 CHI Conference on
  Human Factors in Computing Systems}, ser. CHI '20.\hskip 1em plus 0.5em minus
  0.4em\relax New York, NY, USA: Association for Computing Machinery, 2020, pp.
  1--13. [Online]. Available: \url{https://doi.org/10.1145/3313831.3376191}
\BIBentrySTDinterwordspacing

\bibitem{Xenopoulos2022}
\BIBentryALTinterwordspacing
P.~Xenopoulos, J.~a. Rulff, and C.~Silva, ``Ggviz: Accelerating large-scale
  esports game analysis,'' \emph{Proc. ACM Hum.-Comput. Interact.}, vol.~6, no.
  CHI PLAY, 10 2022. [Online]. Available: \url{https://doi.org/10.1145/3549501}
\BIBentrySTDinterwordspacing

\bibitem{Feitosa2015}
\BIBentryALTinterwordspacing
V.~R.~M. Feitosa, J.~G.~R. Maia, L.~O. Moreira, and G.~A.~M. Gomes, ``Gamevis:
  Game data visualization for the web,'' in \emph{2015 14th Brazilian Symposium
  on Computer Games and Digital Entertainment (SBGames)}, 2015, pp. 70--79.
  [Online]. Available: \url{https://doi.org/10.1109/SBGames.2015.21}
\BIBentrySTDinterwordspacing

\bibitem{Afonso2019}
A.~P. Afonso, M.~B. Carmo, and T.~Moucho, ``Comparison of visualization tools
  for matches analysis of a moba game,'' in \emph{2019 23rd International
  Conference Information Visualisation (IV)}, 2019, pp. 118--126.

\bibitem{Kuan2017}
\BIBentryALTinterwordspacing
Y.-T. Kuan, Y.-S. Wang, and J.-H. Chuang, ``Visualizing real-time strategy
  games: The example of starcraft ii,'' in \emph{2017 IEEE Conference on Visual
  Analytics Science and Technology (VAST)}, 2017, pp. 71--80. [Online].
  Available: \url{https://doi.org/10.1109/VAST.2017.8585594}
\BIBentrySTDinterwordspacing

\bibitem{Rijnders2022}
\BIBentryALTinterwordspacing
F.~Rijnders, G.~Wallner, and R.~Bernhaupt, ``Live feedback for training through
  real-time data visualizations: A study with league of legends,'' \emph{Proc.
  ACM Hum.-Comput. Interact.}, vol.~6, no. CHI PLAY, oct 2022. [Online].
  Available: \url{https://doi.org/10.1145/3549506}
\BIBentrySTDinterwordspacing

\bibitem{Wallner2021}
\BIBentryALTinterwordspacing
G.~Wallner, M.~van Wijland, R.~Bernhaupt, and S.~Kriglstein, ``What players
  want: Information needs of players on post-game visualizations,'' in
  \emph{Proceedings of the 2021 CHI Conference on Human Factors in Computing
  Systems}, ser. CHI '21.\hskip 1em plus 0.5em minus 0.4em\relax New York, NY,
  USA: Association for Computing Machinery, 2021. [Online]. Available:
  \url{https://doi.org/10.1145/3411764.3445174}
\BIBentrySTDinterwordspacing

\bibitem{Kowalczyk2018}
\BIBentryALTinterwordspacing
N.~Kowalczyk, F.~Shi, M.~Magnuski, M.~Skorko, P.~Dobrowolski, B.~Kossowski,
  A.~Marchewka, M.~Bielecki, M.~Kossut, and A.~Brzezicka, ``Real-time strategy
  video game experience and structural connectivity – a diffusion tensor
  imaging study,'' \emph{Human Brain Mapping}, vol.~39, no.~9, pp. 3742--3758,
  2018. [Online]. Available: \url{https://doi.org/10.1002/hbm.24208}
\BIBentrySTDinterwordspacing

\bibitem{Chen2021}
\BIBentryALTinterwordspacing
M.~A. Chen, K.~Spanton, P.~van Schaik, I.~Spears, and D.~Eaves, ``{The Effects
  of Biofeedback on Performance and Technique of the Boxing Jab},''
  \emph{Perceptual and Motor Skills}, vol. 128, no.~4, pp. 1607--1622, 2021,
  pMID: 33940988. [Online]. Available:
  \url{https://doi.org/10.1177/00315125211013251}
\BIBentrySTDinterwordspacing

\bibitem{Lapi2018}
\BIBentryALTinterwordspacing
O.~Lappi, ``The racer's mind—how core perceptual-cognitive expertise is
  reflected in deliberate practice procedures in professional motorsport,''
  \emph{Frontiers in Psychology}, vol.~9, 2018. [Online]. Available:
  \url{https://doi.org/10.3389/fpsyg.2018.01294}
\BIBentrySTDinterwordspacing

\bibitem{Perrey2022}
\BIBentryALTinterwordspacing
S.~Perrey, ``Training monitoring in sports: It is time to embrace cognitive
  demand,'' \emph{Sports}, vol.~10, no.~4, 2022. [Online]. Available:
  \url{https://doi.org/10.3390/sports10040056}
\BIBentrySTDinterwordspacing

\bibitem{Walton2018}
\BIBentryALTinterwordspacing
C.~C. Walton, R.~J. Keegan, M.~Martin, and H.~Hallock, ``The potential role for
  cognitive training in sport: More research needed,'' \emph{Frontiers in
  Psychology}, vol.~9, 2018. [Online]. Available:
  \url{https://doi.org/10.3389/fpsyg.2018.01121}
\BIBentrySTDinterwordspacing

\bibitem{Kelly2021}
\BIBentryALTinterwordspacing
S.~Kelly and J.~Leung, ``The new frontier of esports and gaming: A scoping
  meta-review of health impacts and research agenda,'' \emph{Frontiers in
  Sports and Active Living}, vol.~3, 2021. [Online]. Available:
  \url{https://doi.org/10.3389/fspor.2021.640362}
\BIBentrySTDinterwordspacing

\bibitem{Banyai2019}
\BIBentryALTinterwordspacing
F.~B{\'a}nyai, M.~D. Griffiths, O.~Kir{\'a}ly, and Z.~Demetrovics, ``The
  psychology of esports: A systematic literature review,'' \emph{Journal of
  Gambling Studies}, vol.~35, no.~2, pp. 351--365, Jun 2019. [Online].
  Available: \url{https://doi.org/10.1007/s10899-018-9763-1}
\BIBentrySTDinterwordspacing

\bibitem{Voisin2022}
\BIBentryALTinterwordspacing
N.~Voisin, N.~Besombes, and S.~Laffage-Cosnier, ``Are esports players inactive?
  a systematic review,'' \emph{Physical Culture and Sport. Studies and
  Research}, vol.~97, no.~1, pp. 32--52, 2022. [Online]. Available:
  \url{https://doi.org/10.2478/pcssr-2022-0022}
\BIBentrySTDinterwordspacing

\bibitem{Sabtan2022}
\BIBentryALTinterwordspacing
B.~Sabtan, S.~Cao, and N.~Paul, ``Current practice and challenges in coaching
  esports players: An interview study with league of legends professional team
  coaches,'' \emph{Entertainment Computing}, vol.~42, p. 100481, 2022.
  [Online]. Available: \url{https://doi.org/10.1016/j.entcom.2022.100481}
\BIBentrySTDinterwordspacing

\bibitem{Poulus2022}
\BIBentryALTinterwordspacing
D.~R. Poulus, T.~J. Coulter, M.~G. Trotter, and R.~Polman, ``A qualitative
  analysis of the perceived determinants of success in elite esports
  athletes,'' \emph{Journal of Sports Sciences}, vol.~40, no.~7, pp. 742--753,
  2022, pMID: 34930102. [Online]. Available:
  \url{https://doi.org/10.1080/02640414.2021.2015916}
\BIBentrySTDinterwordspacing

\bibitem{Kari2016}
\BIBentryALTinterwordspacing
T.~Kari and V.-M. Karhulahti, ``Do e-athletes move?'' \emph{International
  Journal of Gaming and Computer-Mediated Simulations}, vol.~8, no.~4, pp.
  53--66, oct 2016. [Online]. Available:
  \url{https://doi.org/10.4018/ijgcms.2016100104}
\BIBentrySTDinterwordspacing

\bibitem{Fanfarelli2018}
\BIBentryALTinterwordspacing
J.~R. Fanfarelli, ``Expertise in professional overwatch play,''
  \emph{International Journal of Gaming and Computer-Mediated Simulations
  (IJGCMS)}, vol.~10, no.~1, pp. 1--22, 2018. [Online]. Available:
  \url{https://doi.org/10.4018/IJGCMS.2018010101}
\BIBentrySTDinterwordspacing

\bibitem{Abbott2023}
\BIBentryALTinterwordspacing
C.~Abbott, M.~Watson, and P.~Birch, ``Perceptions of effective training
  practices in league of legends: A qualitative exploration,'' \emph{Journal of
  Electronic Gaming and Esports}, vol.~1, no.~1, 2023. [Online]. Available:
  \url{https://doi.org/10.1123/jege.2022-0011}
\BIBentrySTDinterwordspacing

\bibitem{Pereira2022}
\BIBentryALTinterwordspacing
A.~Monteiro~Pereira, J.~A. Costa, E.~Verhagen, P.~Figueiredo, and J.~Brito,
  ``Associations between esports participation and health: A scoping review,''
  \emph{Sports Medicine}, vol.~52, no.~9, pp. 2039--2060, Sep 2022. [Online].
  Available: \url{https://doi.org/10.1007/s40279-022-01684-1}
\BIBentrySTDinterwordspacing

\bibitem{Schubert2022}
\BIBentryALTinterwordspacing
M.~Schubert, F.~Eing, and T.~Könecke, ``Perceptions of professional esports
  players on performance-enhancing substances,'' \emph{Performance Enhancement
  \& Health}, vol.~10, no.~4, p. 100236, 2022. [Online]. Available:
  \url{https://doi.org/10.1016/j.peh.2022.100236}
\BIBentrySTDinterwordspacing

\bibitem{Radford2022Whisper}
\BIBentryALTinterwordspacing
A.~Radford, J.~W. Kim, T.~Xu, G.~Brockman, C.~McLeavey, and I.~Sutskever,
  ``Robust speech recognition via large-scale weak supervision,'' 2022.
  [Online]. Available: \url{https://arxiv.org/abs/2212.04356}
\BIBentrySTDinterwordspacing

\bibitem{DiFrancisco2021}
\BIBentryALTinterwordspacing
J.~DiFrancisco-Donoghue, S.~E. Jenny, P.~C. Douris, S.~Ahmad, K.~Yuen,
  T.~Hassan, H.~Gan, K.~Abraham, and A.~Sousa, ``{Breaking up prolonged sitting
  with a 6 min walk improves executive function in women and men esports
  players: a randomised trial},'' \emph{BMJ Open Sport \& Exercise Medicine},
  vol.~7, no.~3, 2021. [Online]. Available:
  \url{http://dx.doi.org/10.1136/bmjsem-2021-001118}
\BIBentrySTDinterwordspacing

\bibitem{Mandolesi2018}
\BIBentryALTinterwordspacing
L.~Mandolesi, A.~Polverino, S.~Montuori, F.~Foti, G.~Ferraioli, P.~Sorrentino,
  and G.~Sorrentino, ``Effects of physical exercise on cognitive functioning
  and wellbeing: Biological and psychological benefits,'' \emph{Frontiers in
  Psychology}, vol.~9, 2018. [Online]. Available:
  \url{https://doi.org/10.3389/fpsyg.2018.00509}
\BIBentrySTDinterwordspacing

\bibitem{Brown2021}
\BIBentryALTinterwordspacing
B.~M. Brown, N.~Frost, S.~R. Rainey-Smith, J.~Doecke, S.~Markovic, N.~Gordon,
  M.~Weinborn, H.~R. Sohrabi, S.~M. Laws, R.~N. Martins, K.~I. Erickson, and
  J.~J. Peiffer, ``High-intensity exercise and cognitive function in
  cognitively normal older adults: a pilot randomised clinical trial,''
  \emph{Alzheimer's Research {\&} Therapy}, vol.~13, no.~1, p.~33, Feb 2021.
  [Online]. Available: \url{https://doi.org/10.1186/s13195-021-00774-y}
\BIBentrySTDinterwordspacing

\bibitem{Spitzer2022}
\BIBentryALTinterwordspacing
P.~Spitzer, N.~Kühl, and M.~Goutier, ``Training novices: The role of human-ai
  collaboration and knowledge transfer,'' 2022. [Online]. Available:
  \url{https://arxiv.org/abs/2207.00497}
\BIBentrySTDinterwordspacing

\end{thebibliography}

\end{document}